# Do ResearchGate Scores create ghost academic reputations?

Enrique Orduna-Malea[1], Alberto Martín-Martín[2], Mike Thelwall[3] and Emilio Delgado López-Cózar[2*]

[1] *EC3 Research Group, Polytechnic University of Valencia. Camino de Vera s/n, Valencia 46022, Spain*
[2] *EC3 Research Group, Universidad de Granada, 18071 Granada, Spain*
[3] *Statistical Cybermetrics Research Group, School of Mathematics and Computer Science. University of Wolverhampton, Wulfruna Street, Wolverhampton WV1 1LY, UK*

**\***edelgado@ugr.es

**Abstract:** The academic social network site ResearchGate (RG) has its own indicator, RG Score, for its members. The high profile nature of the site means that the RG score may be used for recruitment, promotion and other tasks for which researchers are evaluated. In response, this study investigates whether it is reasonable to employ the RG Score as evidence of scholarly reputation. For this, three different author samples were investigated. An outlier sample includes 104 authors with high values. A Nobel sample comprises 73 Nobel winners from Medicine & Physiology, Chemistry, Physics and Economics (from 1975 to 2015). A longitudinal sample includes weekly data on 4 authors with different RG Scores. The results suggest that high RG Scores are built primarily from activity related to asking and answering questions in the site. In particular, it seems impossible to get a high RG Score solely through publications. Within RG it is possible to distinguish between (passive) academics that interact little in the site and active platform users, who can get high RG Scores through engaging with others inside the site (questions, answers, social networks with influential researchers). Thus, RG Scores should not be mistaken for academic reputation indicators.

**Acknowledgments:** Alberto Martín-Martín enjoys a four-year doctoral fellowship (FPU2013/05863) granted by the Ministerio de Educación, Cultura, y Deporte (Spain). Enrique Orduna-Malea holds a postdoctoral fellowship (PAID-10-14), from the Polytechnic University of Valencia (Spain).

## 1. Introduction

The use of bibliometric indicators in academic decision-making (e.g., funding, accreditation, tenure, promotion or recruitment), has produced some negative and pernicious effects. Science policy-makers may be tempted to endorse bibliometric indicators to simplify the hard task of evaluating performance (Jiménez-Contreras, de Moya Anegon and Delgado López-Cózar 2003). The value of this approach is currently being debated, with guidelines being proposed about the type of indicators that would be useful in different contexts (Wilsdon, Allen, Belfiore, Campbell, Curry, et al. 2015).

Online academic profiles, such as AMiner[1], Microsoft Academic Search[2], Google Scholar Citations[3], ResearcherID[4], ORCID[5] and academic social networks like Mendeley[6], Academia.edu[7] and ResearchGate[8] have become an accepted part of the academic landscape (Martín-Martín, Orduna-Malea and Delgado López-Cózar 2016; Ortega 2016). Some provide a range of metrics for both authors and articles (Orduna-Malea, Martín-Martín and Delgado López-Cózar 2016) and, assuming that these metrics are used, it is important to understand their key properties.

RG has become one of the most used online academic social tools (Van Noorden 2014; Bosman and Kramer 2016), although with disciplinary differences in uptake and patterns of use (Jordan 2014a). Its main indicator is the RG Score, which is displayed prominently on author profile pages but is not defined in the site. The popularity of RG and the feasibility of directly including it on an automatic exportable résumé built from the user's information in the platform, suggest that it could be used by evaluators in recruiting tasks for research positions or grants, especially if they use the RG job vacancies service for academics[9].

Most academic studies of RG have focused on its features and social functions or its degree of use in specific fields. There are few critical evaluations of RG indicators (mainly the RG score). The two main exceptions (Kraker and Lex 2015; Jordan 2015) are discussed below, and both mainly focus on academics with low RG scores. There is thus a danger that the RG score may be employed in evaluative tasks without a full understanding of its properties. In response, the objective of this work is to better understand the RG Score in order to assess whether it is reasonable to use it as an academic reputation indicator.

## 2. Background

ResearchGate was created in Germany in 2008, and by November 2016 claimed to have 11 million users and 100 million publications[10], of which 25% are open access (RG no longer reports the number of open access documents). Its members have a profile page that can list their scientific contributions (whether published or not), co-authors and basic professional information. Authors can be aggregated at the university and entity (Department, Faculty, School, Research group, etc.) levels based upon their self-reported affiliation. The metrics added to user profiles by RG include the number of visits, downloads, reads and citations received. These indicators cannot be taken at face value, however, since they may be spammed or represent automated accesses. RG is also a social network service because it allows members to connect to each other by following them. In addition, it encourages discussions, communities and questions in order to support interactions between members.

A range of studies have investigated the RG interface or discussed the site in general terms (Goodwin, Jeng and He 2014; Hoffmann, Lutz and Meckel 2015; Jordan 2014b; Kadriu 2013; Li et al. 2015; Ovadia 2014; Matthews 2016). Others have sought to identify communities of researchers within the system, such as Spanish university teachers in the area of Communication (González-Díaz, Iglesias-García and Codina 2015), finding weak presences. One recent paper has also argued that the presence of poor quality journals undermines the value of the RG Score (Memon, 2016).

Some studies have investigated the relationship between the indicators reported by RG and bibliometric indicators from traditional bibliographic databases. This is the case of Ortega (2015) about the researchers belonging to the Spanish National Research Council (CSIC), Mikki et al (2016) on the researchers of the University of Bergen, and finally Martin-Martin et al (2016) on the international bibliometric community. Despite very different samples, these three works find low correlations between traditional bibliometric (citation-based) and altmetric (social activity-based) indicators at the author level.

Beyond the direct analysis of authors, other works have focused on demographic aspects related to other units, such as universities or articles. In the case of universities, ResearchGate statistics correlate well with other academic institution rankings, broadly reflecting thus traditional academic capital (Thelwall and Kousha, 2015). In the case of articles, there is uneven coverage according to disciplines, and a low to moderate correlation between view counts and Scopus citations (Thelwall and Kousha, 2017). This may convert ResearchGate Reads in a new audience indicator in their own right.

### The RG Score

The RG Score is claimed to "measure scientific reputation based on how all of your research is received by your peers"[11]. The three components taken into account are: contributions uploaded by the author (e.g., articles, presentations, reports, working papers, raw data); interactions with other members; and reputation gained from other researchers. For the interaction component, engagements with high RG Score members have higher weightings. As a result of the RG Score incorporating site-specific interaction data, its values can differ from indicators that rely solely upon publications (Orduna-Malea, Martín-Martín and Delgado López-Cózar 2016).

The exact composition of these three components is unknown, as is their weighting, but clicking on an individual score gives a breakdown of its origin. From this information, it seems that the RG Score has four dimensions: *Publications*; *Answers*; *Questions*; and *Followers*. For example, an author's RG Score might be decomposed as follows: *Publications*: 50%; *Answers*: 25%; *Questions*: 24%; *Followers*: 1%. Nevertheless, the exact formula for each of these components is unknown, as is the method of combining them.

The RG Score has important deficiencies that seem to prevent it from being used as a scientific reputation measure (Kraker and Lex 2015): a) it is not transparent or reproducible; b) it may incorporate Journal Impact Factors (JIFs) which have well known limitations (it is not clear that it still includes JIFs because it no longer reports them within a separate Impact Points indicator); and c) it has been modified repeatedly and so cannot be tracked over time or compared between periods.

One study has attempted to reproduce the RG Score using different author samples (Jordan 2015). The first sample includes 30 users with one publication, no answers or questions, and with less than 1,000 profile views. For these authors, their RG Score can be predicted with an apparently high level of accuracy (no details are given) by a linear formula using the log of the impact points of their publication (i.e. JIF). Thus, for single paper academics without questions or answers, their RG Score is essentially (a transformation of) the log of the JIF of their publication.

The same study fitted a linear regression model to a variety of factors that might affect RG Scores. This used an expanded sample that included 30 academics with multiple publications but no questions and answers, and 30 with multiple questions and multiple answers, for an overall total of 90, all with under 1000 profile views. The factors modelled included the main data reported by RG, as well as its natural log and two way interactions. The linear regression fitting method and information about the data distribution were not reported. After eliminating non-significant factors, the key predictors of the RG Score were: ln(IF), ln(answers), ln(IF)^2, views, publications, and ln(IF)*ln(publications). Although the formula based on these could predict RG Scores with a high degree of accuracy (from a visual inspection of the graph), the presence of some outliers and strange terms in the formula (e.g., ln(IF)*ln(publications)) suggest that it is not the full story. This may be due to the relatively small sample size (90), which risks the statistical problem of overfitting, given the large number of factors tested in the study.

## 3. Research goals

The goal of this research is to empirically test the reliability of the RG Score as a scholarly reputation indicator with larger samples and different types of samples compared to those previously used. The first two questions mirror those of a previous paper (Jordan 2015) but address them with a much larger data set and may therefore yield more comprehensive answers. The final question is important to understand the effect of using an indicator that incorporates social interactions in place of purely scholarly indicators (e.g., citation counts).
- RQ1: Which RG data influences the RG Score?
- RQ2: Can RG Scores be estimated from RG data?
- RQ3: How strongly do the academic related metrics (contributions, citations, h-index) relate to the RG social connectivity metrics (followers, questions, answers).

## 4. Methods

Three non-random, purposive samples of authors with a public ResearchGate profile were gathered as the raw data for the study[12], retrieving a range of parameters for each (Table 1).

**Table 1. ResearchGate metrics and scope (August 2016)**

| METRIC | SCOPE |
|---|---|
| **Publications** | Number of items published by an author in RG |
| **Reads** | Total number of Reads received by an author |
| **Citations** | Total number of citations to the publications published in RG |
| **Profile views** | Number of times that the author profile has been visited |
| **Impact Points** | Summation of Journal Citation Reports' Impact Factor of each journal in which the author has published their publications available in RG |
| **Total H- Index** | An author has an h-index of "h" when at least "h" of his/her publications achieve at least "h" citations each. |
| **Pure H-Index** | H-index is calculated extracting the self-citations |
| **Following** | Number of RG users that one author is following |
| **Followers** | Number of RG users who follow one author |
| **Answers** | Number of answers performed by an author in RG |
| **Questions** | Number of questions delivered to RG by the author |
| **RG Score** | Composite indicator purposed to measure users' scientific reputation |
| **RG Score – Publication dimension** | Percentage contribution of "Publication" dimension to the total RG Score. |
| **RG Score – Answer dimension** | Percentage contribution of "Answer" dimension to the total RG Score. |
| **RG Score – Question dimension** | Percentage contribution of "Question" dimension to the total RG Score. |
| **RG Score – Followers dimension** | Percentage contribution of "Followers" dimension to the total RG Score. |

*Outlier sample* This sample consists of 104 authors with high values in the global RG Score or any of the main metrics in Table 1. The author gathering process started from a core list of RG members with a RG Score over 100 points[13]. After this, we browsed their following and follower authors, manually extracting all RG users surpassing 100 points, repeating the process iteratively until we stopped identifying new authors with more than 100 points. Next, for each author we identified high valued metrics. ResearchGate's advanced search feature was used to identify users with specific levels of performance in each of the author-level metrics, but this procedure has since been withdrawn by ResearchGate.

It is not possible to assess the comprehensiveness of this sample because ResearchGate does not provide a master list of users, but it seems likely that a high percentage of authors with a RG Score above 100 have been identified, as well as the authors with the highest individual scores (citations, followers, answers, etc.). All authors and metrics are available in the supplementary material (Appendix A).

*Nobel sample* Nobel Prize winners form a useful gold standard of research excellence. ResearchGate claimed in June 2016 to include 52 Nobel Prize winners with a public profile. Nobel Prize winner profiles are characterized by a special badge and details. Other awards, such as the Wolf Medal or the Fields Medal are displayed in a similar fashion. Nobel Prize winners were identified from the official website[14] for all scientists in the Medicine & Physiology, Chemistry, Physics and Economic sciences from 1975 to 2015. Each winner was subsequently searched for in ResearchGate through the basic author search feature. Different name variants were used when needed. This manual process identified 46 (out of 52) RG profiles with Nobel Prize winner badges. An additional set of 26 Nobel Laureates not identified as such by ResearchGate was also found, giving a final sample of 73 authors (see Appendix B). The Peace and Literature prizes were not considered since they may not have scientific contributions. For example, no winners from the last 5 years could be found on RG.

*Longitudinal sample* Weekly scores were gathered from 4 authors with different RG Scores and academic statuses (from a PhD student to a full Professor) for 6 weeks in May and June 2016: RG Score, Impact points, Reads, Citations, Publications, Profile views (Appendix D).

*Analyses*
Descriptive statistics were used to identify evidence about how different contributions might affect RG Scores. Spearman correlations were calculated between the RG profile data for the outlier and Nobel samples in order to identify relationships between the components of the score. Spearman correlations were used due to the skewed data distributions.

A statistical model to estimate the RG Score was fitted to the outlier and Nobel samples. Non-linear regression models were used due to the skewness of web and citation data. Since the purpose was to fit a set of "m" observations (96 outliers and 65 Nobels) with a model that is non-linear in "n" unknown parameters (m > n), a non-linear least squares fitting method was used. The XLStat statistical suite provides a set of built-in non-linear functions operating under the method, and all of them were tested in this work. The Levenberg-Marquardt algorithm was used to fit non-linear regression models to data with one independent variable (RG Score) and eleven dependent variables (publications, reads, citations, profile views, Impact Points, h-index, h-index without citations, followings, followers, questions and answers). All variables consist exclusively of quantitative data (an assumption for fitting). As a measure of model validity, both the coefficient of determination ($R^2$) and scatter plots of residuals versus predictors were used (supplementary material).

Since the RG Score is presumably calculated by a human-designed algorithm and may be designed to be not straightforward so that it is not easily guessed, it is not possible to use the traditional approach to statistical model fitting by using theory to select a range of models to fit. Instead, an ad-hoc approach was taken by trying a wide range of different types of formulae in order to get insights into how the RG Score algorithm might work. Since there is a limited amount of data, a range of different models was tested, and the models have many parameters that are chosen during the fitting process, it is almost inevitable that some models fit well even if they are completely unrelated to the real RG Score approach. Hence, the outcome of the model fitting experiment cannot give valid statistical evidence about the approach used by ResearchGate but can only give insights into how a model might work.

## 5. Results

*Outlier sample*

From the 26 academics with a RG Score over 100 (Table 2), 25 have scores that are primarily from the *Answers* category and two (Pimiskern and Tsambani) have scores exclusively from Answers. In contrast, one author's (Eidiani) high score is dominated (48%) by the *Questions* category.

**Table 2. Authors in the outlier sample with RG Score > 100**

| R | AUTHOR | RG Score | Publications (%) | Answers (%) | Questions (%) | Followers (%) |
|---|---|---|---|---|---|---|
| 1 | Shapiro, Adam B. | 439.82 | 8 | 92 | 0 | 0 |
| 2 | Ewalds-Kvist, Béatrice Marianne | 289.82 | 9 | 90 | 1 | 0 |
| 3 | Karaman, Rafik | 254.11 | 16 | 82 | 2 | 0 |
| 4 | Pimiskern, Joachim | 215.03 | 0 | 100 | 0 | 0 |
| 5 | Liger, Dominique | 181.39 | 15 | 85 | 0 | 0 |
| 6 | Celzard, Alain | 173.99 | 25 | 69 | 6 | 0 |
| 7 | Umachandran, Krishnan | 170.69 | 3 | 94 | 3 | 0 |
| 8 | Kennedy, Ian | 154.65 | 6 | 92 | 2 | 0 |
| 9 | Whitehead, Dean | 153.99 | 22 | 69 | 9 | 0 |
| 10 | Lemkul, Justin | 151.75 | 19 | 81 | 0 | 0 |
| 11 | Muss, Wolfgang H. | 151.63 | 22 | 72 | 6 | 0 |
| 12 | Galllup, Jack | 149.32 | 22 | 78 | 0 | 0 |
| 13 | Brender, Jeffrey | 141.29 | 26 | 64 | 10 | 0 |
| 14 | Brassard, Louis | 137.22 | 2 | 92 | 6 | 0 |
| 15 | Krieger, Hanno | 136.91 | 17 | 82 | 1 | 0 |
| 16 | Peter, James | 132.95 | 28 | 62 | 10 | 0 |
| 17 | Björnsson, Björn Thrandur | 124.42 | 33 | 43 | 24 | 0 |
| 18 | Burzynski, Artur | 123.37 | 23 | 73 | 4 | 0 |
| 19 | Banhegyi, Gyorgy | 123.09 | 23 | 72 | 5 | 0 |
| 20 | Jacić, Ljubomir | 115.02 | 16 | 64 | 20 | 0 |
| 21 | Pasternak, Taras | 112.93 | 25 | 75 | 0 | 0 |
| 22 | Chartrand, Max Stanley | 112.83 | 20 | 72 | 8 | 0 |
| 23 | Eidiani, Mostafa | 110.69 | 14 | 38 | 48 | 0 |
| 24 | Tsambani, Ariadne | 109.8 | 0 | 100 | 0 | 0 |
| 25 | Moss, Marcia | 104.86 | 34 | 64 | 2 | 0 |
| 26 | Morales Pedraza, Jorge | 103.81 | 26 | 70 | 4 | 0 |

The top outlier author with *Publications* as their main RG category is 28[th] (Enzo, RG Score: 92.73). Most (40 out 57) of these academics have *Publications* accounting for 100% of their RG Score, whereas only 2 out of the 47 have contributed any *Answers*. Only one outlier author has a RG Score based on *Questions*, and none on *Followers*.

The *Followers* dimension is almost irrelevant for the outlier category because only two authors (Ebrahim: 1%; Repiso: 3%) have a non-zero score in this dimension. Whilst this could have been due to limitations of the browsing method used to find the sample, the low contributions of Followers to RG Scores suggests that this component is not powerful enough to generate high RG Scores on its own. Overall, *Answers* and *Publications* are the main activities for outlier authors, followed by *Questions* (Figure 1).

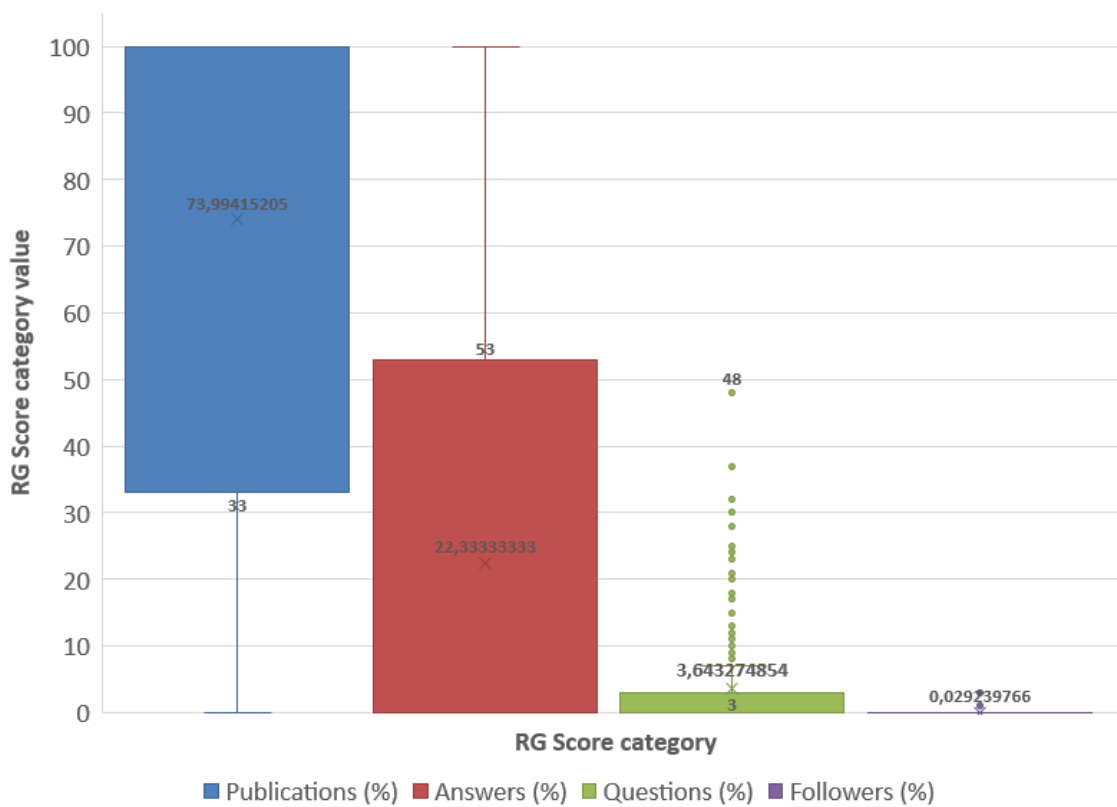

**Figure 1. Outlier author profiles by RG score category.**

The percentages reported by RG do not seem to follow simple relationship with the corresponding metrics. For example, Repiso's 1,152 followers give a 3% *Following* category score, but Kotsemir's 4,737 followers give a 0% *Following* category score. Similarly, despite Shapiro's 2,914 citations and 213.43 Impact Points, his *Publications* category provides only 9% of his huge RG score (439.82), presumably because of his active question answering (2,889 answers). Moreover, not all answers have the same worth since their value may be affected by votes from other users and perhaps also the reputations of these users.

The RG Scores for outlier authors have little association with citation-based metrics, such as publications (r= -0.21), citations (r= -0.09), Impact Points (r= -0.07) or h-index (r= -0.08) (Table 3). The strongest correlations with RG Scores are for Answers (r= 0.61), and Profile Views (r= 0.42).

Table 3. Correlations between RG metrics (outlier authors; n= 104)

|  | Publications | Reads | Citations | Profile views | Impact Points | Total H-index | Pure Hindex | Following | Followers | Questions | Answers | RG Score |
|---|---|---|---|---|---|---|---|---|---|---|---|---|
| Publications | 1.00 | **0.74 | **0.85 | **-0.28 | **0.80 | **0.84 | **0.81 | **-0.34 | -0.11 | **-0.51 | **-0.67 | -0.21 |
| Reads | **0.74 | 1.00 | **0.58 | 0.20 | **0.43 | **0.58 | **0.53 | -0.04 | 0.20 | -0.17 | **-0.37 | -0.12 |
| Citations | **0.85 | **0.58 | 1.00 | **-0.41 | **0.91 | **1.00 | **0.99 | **-0.51 | -0.26 | **-0.68 | **-0.72 | -0.09 |
| Profile views | **-0.28 | 0.20 | **-0.41 | 1.00 | **-0.51 | **-0.40 | **-0.43 | **0.61 | **0.72 | **0.64 | **0.70 | **0.42 |
| ImpactPoints | **0.80 | **0.43 | **0.91 | **-0.51 | 1.00 | **0.90 | **0.92 | **-0.58 | **-0.35 | **-0.73 | **-0.74 | -0.07 |
| Total H-index | **0.84 | **0.58 | **1.00 | **-0.40 | **0.90 | 1.00 | **0.99 | **-0.50 | -0.25 | **-0.67 | **-0.71 | -0.08 |
| Pure Hindex | **0.81 | **0.53 | **0.99 | **-0.43 | **0.92 | **0.99 | 1.00 | **-0.53 | **-0.26 | **-0.69 | **-0.70 | -0.05 |
| Following | **-0.34 | -0.04 | **-0.51 | **0.61 | **-0.58 | **-0.50 | **-0.53 | 1.00 | **0.71 | **0.72 | **0.60 | 0.14 |
| Followers | -0.11 | 0.20 | -0.26 | **0.72 | **-0.35 | -0.25 | **-0.26 | **0.71 | 1.00 | **0.54 | **0.50 | 0.22 |
| Questions | **-0.51 | -0.17 | **-0.68 | **0.64 | **-0.73 | **-0.67 | **-0.69 | **0.72 | **0.54 | 1.00 | **0.78 | 0.26 |
| Answers | **-0.67 | **-0.37 | **-0.72 | **0.70 | **-0.74 | **-0.71 | **-0.70 | **0.60 | **0.50 | **0.78 | 1.00 | **0.61 |
| RG Score | -0.21 | -0.12 | -0.09 | **0.42 | -0.07 | -0.08 | -0.05 | 0.14 | 0.22 | 0.26 | **0.61 | 1.00 |

** Significant values (except diagonal) at the level of significance alpha=0.010 (two-tailed test).

We tested a set of non-linear regression models for RG Score (see Appendix A). The best model obtained a coefficient of determination equal to 0.68. Figure 2 displays the estimated values for this model against the real RG Scores and the estimated values provided by the Jordan simplified model (2015), mentioned previously (Y= 1.562 Ln(x) + 1.5878). The Jordan extended model cannot be directly applied as it utilizes the number of views, a metric no longer available in ResearchGate.

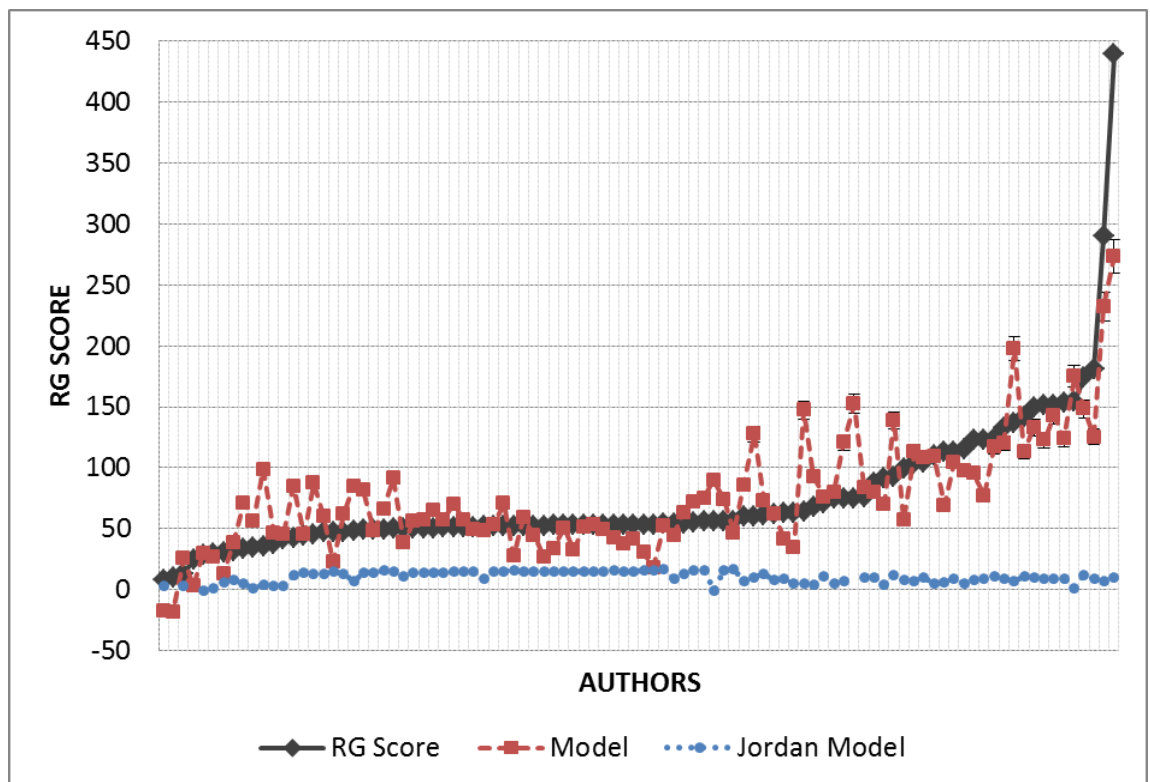

**Figure 2. RG Score estimates using non-linear regression models (outlier authors)**

The correlation between the real RG Score and the model estimates is high (r= 0.83). Nonetheless, the estimates are not reliable. In only 16 out of 104 observations (15.4%), is the residual value (difference between the real and estimated value) lower than 5 points. The Jordan model (which estimates RG score uniquely from Impact Points) works less well, although it was conceived to estimate low author scores. The correlation between the real and Jordan estimated values is small (r= -0.13).

*Nobel sample*

The 73 Nobel winners with a public RG profile include 67 with a RG Score. All construct their RG Score mainly from the *Publications* category (e.g., Table 4) and 64 have scores that are 100% in this category. The three minor exceptions are Gurdon (1% in *Answers*), Hooft (20% in *Answers* and 3% in *Questions*), and Stiglitz (1% in *Followers*).

**Table 4. The 25 Nobel authors with the highest RG Scores**

| N | Name | Area | RG Score | Publications (%) | Answers (%) | Questions (%) | Followers (%) |
|---|------|------|----------|------------------|-------------|---------------|---------------|
| 1 | Greengard, Paul | Physiology or Medicine | **54.18** | 100 | 0 | 0 | 0 |
| 2 | Perl, Martin L | Physics | **52.12** | 100 | 0 | 0 | 0 |
| 3 | Grubbs, Robert H | Chemistry | **52.09** | 100 | 0 | 0 | 0 |
| 4 | Wuethrich, Kurt | Chemistry | **51.81** | 100 | 0 | 0 | 0 |
| 5 | Hänsch, Theodor | Physics | **49.55** | 100 | 0 | 0 | 0 |
| 6 | Beutler, Bruce | Physiology or Medicine | **49.44** | 100 | 0 | 0 | 0 |
| 7 | Gurdon, John B | Physiology or Medicine | **48.28** | 99 | 1 | 0 | 0 |
| 8 | Szostak, Jack | Physiology or Medicine | **48.00** | 100 | 0 | 0 | 0 |
| 9 | Kroto, Harold | Chemistry | **47.89** | 100 | 0 | 0 | 0 |
| 10 | Schrock, Richard | Chemistry | **47.82** | 100 | 0 | 0 | 0 |
| 11 | Walker, John E. | Chemistry | **47.73** | 100 | 0 | 0 | 0 |
| 12 | Amano, Hiroshi | Physics | **47.71** | 100 | 0 | 0 | 0 |
| 13 | Moerner, William E. | Chemistry | **47.61** | 100 | 0 | 0 | 0 |
| 14 | Neher, Erwin | Physiology or Medicine | **47.46** | 100 | 0 | 0 | 0 |
| 15 | Guillemin, Roger | Physiology or Medicine | **47.32** | 100 | 0 | 0 | 0 |
| 16 | Crutzen, Paul | Chemistry | **47.02** | 100 | 0 | 0 | 0 |
| 17 | Hooft, Gerard T. | Physics | **46.67** | 77 | 20 | 3 | 0 |
| 18 | Smoot, George | Physics | **46.63** | 100 | 0 | 0 | 0 |
| 19 | Marcus, Rudolph | Chemistry | **45.62** | 100 | 0 | 0 | 0 |
| 20 | Hunt, Tim | Physiology or Medicine | **45.53** | 100 | 0 | 0 | 0 |
| 21 | Levitt, Michael | Chemistry | **45.25** | 100 | 0 | 0 | 0 |
| 22 | Modrich, Paul | Chemistry | **45.22** | 100 | 0 | 0 | 0 |
| 23 | Lindahl, Tomas | Chemistry | **44.69** | 100 | 0 | 0 | 0 |
| 24 | Kobilka, Brian K. | Chemistry | **44.62** | 100 | 0 | 0 | 0 |
| 25 | Warshel, Arieh | Chemistry | **44.58** | 100 | 0 | 0 | 0 |

In contrast to the outliers sample, Nobel authors' RG Scores have statistically significant positive correlations with all citation-based metrics (Publications: r= 0.87; Citations: r= 0.68; h-index: r= 0.85; Impact Points: r= 0.95) and the number of Reads (r= 0.68), in contrast to Questions (r= 0.11) and Answers (r= 0.28) (Table 5).

**Table 5. Correlations between RG metrics (Nobel sample; n= 65)**

|  | Publications | Reads | Citations | Profile views | Impact Points | Total H-index | Pure Hindex | Following | Followers | Questions | Answers | RG Score |
|---|---|---|---|---|---|---|---|---|---|---|---|---|
| Publications | 1.00 | **0.77 | **0.74 | **0.35 | **0.78 | **0.86 | **0.85 | 0.27 | **0.37 | -0.05 | 0.20 | **0.87 |
| Reads | **0.77 | 1.00 | **0.75 | **0.51 | **0.64 | **0.76 | **0.75 | **0.38 | **0.69 | -0.07 | 0.16 | **0.68 |
| Citations | **0.74 | **0.75 | 1.00 | **0.36 | **0.64 | **0.91 | **0.91 | 0.17 | **0.55 | 0.02 | 0.07 | **0.68 |
| Profile views | **0.35 | **0.51 | **0.36 | 1.00 | **0.45 | **0.46 | **0.45 | **0.33 | **0.69 | 0.11 | 0.31 | **0.48 |
| ImpactPoints | **0.78 | **0.64 | **0.64 | **0.45 | 1.00 | **0.82 | **0.81 | 0.25 | 0.29 | -0.08 | 0.12 | **0.95 |
| Total H-index | **0.86 | **0.76 | **0.91 | **0.46 | **0.82 | 1.00 | **1.00 | 0.23 | **0.48 | -0.06 | 0.09 | **0.85 |
| Selective Hindex | **0.85 | **0.75 | **0.91 | **0.45 | **0.81 | **1.00 | 1.00 | 0.22 | **0.48 | -0.06 | 0.09 | **0.83 |
| Following | 0.27 | **0.38 | 0.17 | **0.33 | 0.25 | 0.23 | 0.22 | 1.00 | **0.42 | -0.05 | -0.03 | 0.24 |
| Followers | **0.37 | **0.69 | **0.55 | **0.69 | 0.29 | **0.48 | **0.48 | **0.42 | 1.00 | 0.03 | 0.19 | **0.32 |
| Questions | -0.05 | -0.07 | 0.02 | 0.11 | -0.08 | -0.06 | -0.06 | -0.05 | 0.03 | 1.00 | **0.51 | 0.11 |
| Answers | 0.20 | 0.16 | 0.07 | 0.31 | 0.12 | 0.09 | 0.09 | -0.03 | 0.19 | **0.51 | 1.00 | 0.28 |
| RG Score | **0.87 | **0.68 | **0.68 | **0.48 | **0.95 | **0.85 | **0.83 | 0.24 | **0.32 | 0.11 | 0.28 | 1.00 |

** Significant (except diagonal) at alpha=0.010 (two-tailed test).

The estimation of RG Score values from a non-linear regression model is displayed in Figure 3. In this case, the best fitting model relies exclusively on the publications metric, with a coefficient of determination equal to 0.74. More information about the different models tested is available in the supplementary material (see Appendix B).

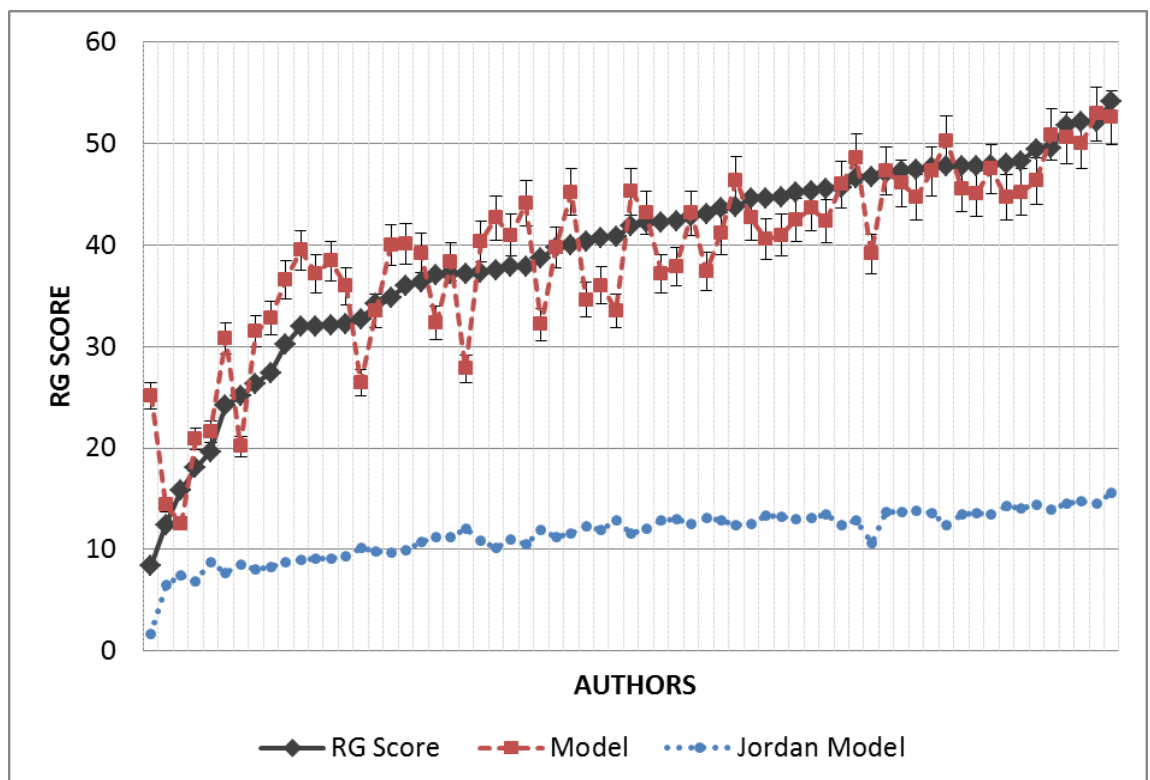

**Figure 3. RG Score estimates using non-linear regression models (Nobel winner authors)**

The estimation of Nobel winners' RG Scores is more accurate than that obtained previously for outliers (Figure 4). In fact, the residual value related to 45 out of the 65 authors is lower than 5 score points. Even the Jordan model - though differing in the raw global value estimated - exhibits a high correlation with the real RG Scores (r= 0.95).

The top ten authors according to each of the RG metrics considered both for the outliers and Nobel samples are available in the supplementary material (Appendix C).

*Longitudinal sample*

The longitudinal sample included authors with the following RG Scores as of June 2016: Author 1: 32.63; Author 2: 24.27; Author 3: 12.35, and Author 4: 10.20 (Table 6, Figure 4). Authors 1 and 3 have publications (7 and 3, respectively) in the period but none in journals included in the Journal Citation Reports, and so do not get impact points for them. Author 2's RG Score decreased (-0.17) due to the elimination of duplicate documents. This occurred despite a significant number of Reads (1,006), Citations (21) and Profile views (456) in the period, confirming the importance of Impact Points for RG scores for authors relying strictly on the publication dimension.

Table 6. RG metric evolution over time (May 1st to June 5th 2016).

| A | RG SCORE | | | | IMPACT POINTS | | | | READS | | | | CITATIONS | | | | PUBLICATIONS | | | | PROFILE VIEWS | | | |
|---|---|---|---|---|---|---|---|---|---|---|---|---|---|---|---|---|---|---|---|---|---|---|---|---|
| | Min | Max | Med | R | Min | Max | Med | R | Min | Max | Med | R | Min | Max | Med | R | Min | Max | Med | R | Min | Max | Med | R |
| 1 | 32.54 | 32.63 | 32.62 | 0.09 | 196.6 | 196.6 | 196.60 | 0 | 20928 | 22432 | 21769 | 1504 | 1116 | 1417 | 1317 | 301 | 241 | 248 | 248 | 7 | 3923 | 4338 | 4222 | 415 |
| 2 | 24.27 | 24.53 | 24.44 | -0.17 | 37.25 | 38.11 | 37.68 | -0.86 | 10290 | 11296 | 10946 | 1006 | 125 | 146 | 136 | 21 | 78 | 82 | 80 | 2 | 1598 | 2054 | 1954 | 456 |
| 3 | 12.09 | 12.35 | 12.23 | 0.24 | 8.99 | 8.99 | 8.99 | 0 | 7980 | 8594 | 8225 | 614 | 37 | 48 | 42 | 10 | 24 | 27 | 25.5 | 3 | 608 | 679 | 659 | 71 |
| 4 | 10.14 | 10.20 | 10.16 | 0.05 | 7.51 | 7.51 | 7.51 | 0 | 7489 | 7798 | 7647 | 309 | 43 | 53 | 49 | 10 | 20 | 20 | 20 | 0 | 286 | 350 | 326 | 64 |

A: Author; Min: Minimun; Max: Maximun; Med: Median; R: Range

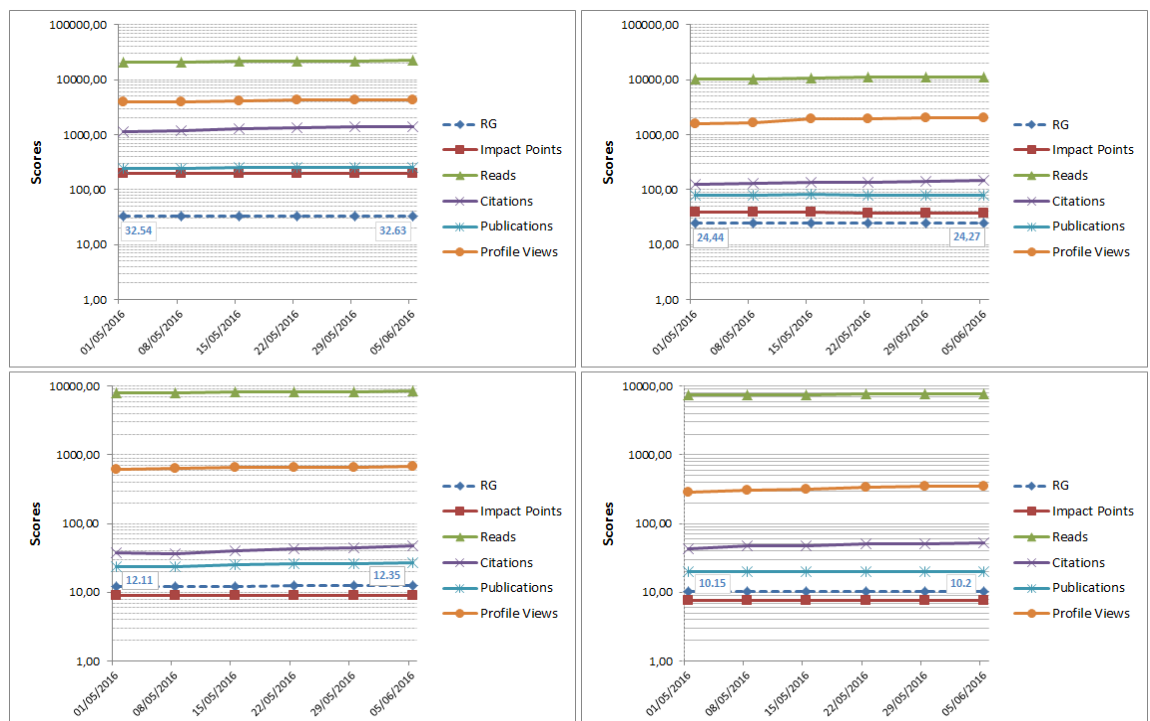

**Figure 4. RG metric evolution (from May 1st to June 5th 2016)**

**6. Discussion and conclusions**

For RQ1, the analyses suggest that the *Answers* dimension is more influential than the remaining categories (*Publications*, *Questions*, and *Followers*). All the high RG Scores identified (over 60 points) are built primarily upon Answers. This explains the high correlation between the RG Score and the number of answers for the outliers and Nobel winners.

Active participation through questions, though important, seems to be less influential. Perhaps the scarcity of new questions compared to answers explains the lower intensity of this parameter. The most questions is 275 (Lala Sukla) whereas most answers is 9,853 (Ljubomir Jacić). Nevertheless, Questions can contribute, as they do for Mostafa Eidiani (53.13 of his 110.69 RG Score from Questions).

For the Nobel winner set that rely mainly on *Publications*, the number of Impact Points dominates RG Scores (see Table 7; see also Table 10), confirming previous findings of Kraker and Lex (2015) with a more comprehensive sample.

As claimed by Kraker and Lex (2015) and Jordan (2015), the relationship between publications and RG Score seems to be logarithmic, making it difficult to achieve a high score from publications alone (see Table 7). The logarithmic relationship may not apply to the other RG categories, however, since one author (Shapiro) has 439.82 points.

Followers seem to have little influence. Even thousands of followers gives can give no increase in RG Score (e.g., Panagiotis Stefanides's 1,261 followers).

For RQ2 RG Scores can be estimated with some accuracy for authors dominated by their Publications category. In this case, the non-linear regression model used provides reasonable but imperfect estimates. The model proposed by Jordan (2015) based on Impact Points works well in this case (r= 0.95).

No effective regression model was found for sets of authors with substantial combinations of Answers, Questions and Following metrics (Figure 3). The RG algorithm is therefore presumably non-linear and perhaps uses additional data or includes weights for some parameters that are conditional on values for others (e.g., questions may only count for people that provide answers). Perhaps also the number of answers submitted by an author is weighted by the positive/negative votes received by them; and the number of followers is adjusted by the RG Scores of these following authors.

The differences obtained between the four Nobel Prize research fields (Chemistry, Medicine, Physics, and Economics) may reflect not only the different citation patterns of each discipline but also their presence on the platform. The lower values for Physics compared to Chemistry may be a consequence of the generalized use of other platforms to deposit physics preprints (e.g., ArXiv). In any case, it is surprising due to the low number of Chemistry articles uploaded to ResearchGate (Thelwall and Kousha 2017). Probably the influence of Biochemistry, and the nature of this sample (Nobel winners) may explain this effect.

The case of Medicine may be due to many users in this field with a RG public profile (Thelwall and Kousha 2017) or high journal Impact Factors. Likewise, although Economics has lower citation scores and Impact Points, the high number of average Reads (comparable with Physics for Nobel winners) confirms the important role of ResearchGate in disseminating social science research results. The number of RG Reads (downloads and views) may be useful for evaluating the media impact and professional influence of contributions.

The results also point to the existence of two different worlds within prominent ResearchGate members. The first (academics) is constituted from authors with many scientific publications and high bibliometric indicators (productivity, citation, and h-index). The second (active RG users) is formed from authors who build their reputation through their communication and collaboration activities within the site.

For active RG users, the RG Score reflects their activities within the site rather than their wider scholarly reputation. This activity generates what Nicholas, Clark and Herman (2016) call reputational anomalies. Nevertheless, RG activity seems likely to be intrinsically positive and beneficial and so it is an open question as to whether peers would regard RG scores based on activities in the site as valid indicators of contributions to research. Nevertheless, RG scores fail the criteria of the Leiden Manifesto (Hicks et al 2015) (Table 7) and so should not be imposed on researchers and should be treated with caution, if used.

**Table 7. RG Score under the Leiden Manifesto\***

| Principle | RG Score | |
|---|---|---|
| | **Full Accomplishment** | **Reason** |
| **1. Quantitative evaluation should support qualitative, expert assessment** | NO | Questions and answers are rated either as positive or negative by users. However, the remaining indicators are exclusively of |
| | | quantitative data |
| **2. Measure performance against the research missions of the institution, group or researcher** | NO | Socio-economic and cultural contexts, and diverse research missions are not taken into account. We acknowledge the identification of different document types, such as teaching and conference material. |
| **3. Protect excellence in locally relevant research** | NO | RG coverage includes locally relevant research. However, only those publications with Impact Factor determine the Impact Point metric, which in turn determines RG Score. |
| **4. Keep data collection and analytical processes open, transparent and simple** | NO | RG Score is not transparent. Both indicators and weights keep under commercial secret. |
| **5. Allow those evaluated to verify data and analysis** | NO | Yes = RG asks users to verify authorship, for example.<br>No = citation metrics and Reads metrics are not verified by authors, being calculated automatically. |
| **6. Account for variation by field in publication and citation practices** | NO | Though RG team indicates they will, currently variations by field are not considered. |
| **7. Base assessment of individual researchers on a qualitative judgement of their portfolio** | NO | Years of scientific career are not considered in h-index calculation.<br>Experience and activities are included in the users' portfolio. However, there is no evidence about the use of this information to build RG Scores.<br>Author's influence is considered through diverse quantitative indicators (such as followers) |
| **8. Avoid misplaced concreteness and false precision** | NO | A great battery of metrics is displayed for each author. However, uncertainty and errors are not signalled. |
| **9. Recognize the systemic effects of assessment and indicators** | NO | Although RG provides a wide battery of indicators, RG Score summarizes authors' reputation. This has led some authors to game the system emphasizing such indicators that influence RG in a greater term (questions and answers) |
| **10. Scrutinize indicators regularly and update them** | NO | RG updates its algorithm several times per year. However, this is not performed under a regular basis, and the platform do not advise users before the update, just only when modifications have been performed. |

\* Source: Hicks et al (2015).

The conclusions are limited by a number of methodological shortcomings. First, the three samples are artificial. The absence of advanced search functions (as well as an official declaration by the ResearchGate team against performing automated queries) makes it difficult to retrieve author rankings according to each of the available metrics. In this context, sampling limitations are almost impossible to avoid.

Second, there is no master list of metrics for each RG Score category. Moreover, the RG Score algorithm changes over time (Kraker and Lex 2015), making long term estimation impossible.

Third, the manipulation of RG Scores by some authors may jeopardize the utilization of this metric for evaluative purposes, even with pure academic users. The lack of filtering makes all RG metrics prone to be gamed.

Finally, whilst this research is critical of the RG Score as an indicator of scholarly reputation, this is not a criticism of the score itself (or their individual metrics, of interest to measure different author dimensions) nor of the general functioning of the ResearchGate platform.

## 7. Notes

1. https://aminer.org
2. http://academic.research.microsoft.com
3. https://scholar.google.com/citations
4. https://www.researcherid.com
5. http://orcid.org
6. https://www.mendeley.com/profiles
7. https://www.academia.edu/
8. https://www.researchgate.net
9. https://www.researchgate.net/jobs
10. https://www.researchgate.net/press
11. https://www.researchgate.net/publicprofile.RGScoreFAQ.html
12. https://doi.org/10.13140/RG.2.2.26322.35526
13. https://www.researchgate.net/post/Which_researcher_has_the_highest_RG_score_and_what_does_that_really_mean
14. http://www.nobelprize.org